\documentclass[twocolumn,showpacs,preprintnumbers,amsmath,amssymb]{revtex4-1}

\usepackage{graphicx}
\def\<{\langle}
\def\>{\rangle}
\def\({\left(}
\def\){\right)}
\def\[{\left[}
\def\]{\right]}
\def\up{\uparrow}
\def\dn{\downarrow}
\def\s{\sigma}
\def\e{\mathrm{e}}
\def\i{\mathrm{i}}
\def\Re#1{\mathrm{Re}\left\{#1\right\}}

\begin{document}
\title{Modeling the Antiferromagnetic Phase in Iron Pnictides: Weakly Ordered State}

\author{E. Kaneshita$^{1}$}
\author{T. Morinari$^1$}
\author{T. Tohyama$^{1,2}$}
\affiliation{
$^1$Yukawa Institute for Theoretical Physics, Kyoto University, Kyoto 606-8502, Japan\\
$^2$JST, Transformative Research-Project on Iron Pnictides (TRIP), Chiyoda, Tokyo 102-0075, Japan}

\date{\today}

\begin{abstract}
We examine electronic states of antiferromagnetic phase in iron pnictides by mean-field calculations of the optical conductivity.
We find that a five-band model exhibiting a small magnetic moment, inconsistent with the first-principles calculations, reproduces well the excitation spectra characterized by a multi-peak structure emerging below the N\'{e}el temperature at low energy, together with an almost temperature-independent structure at high energy.
Investigating the interlayer magnetoresistance for this model, we also predict its characteristic field dependence reflecting the Fermi
surface.
\end{abstract}

\maketitle

\textit{Introduction.}---
Iron pnictide superconductors have received considerable attention since the discovery of superconductivity in LaFeAsO$_{1-x}$F$_{x}$~\cite{kamihara}.
The superconductivity appears next to the antiferromagnetic (AFM) phase that emerges when $x=0$ for LaFeAsO$_{1-x}$F$_{x}$ with the N\'{e}el temperature $T_N\sim137$~K~\cite{Cruz}.
Since the pairing of electrons is expected to be mediated by magnetic fluctuations~\cite{Mazin,Kuroki}, the understanding of the AFM phase is crucial for clarifying the mechanism of the superconductivity in the iron pnictides.

First-principles calculations of the band structure are considered to have enriched our knowledge of the underlying electronic structures in the iron pnictides~\cite{Singh}.
Those calculations, however, disagree with experiments on the magnitude of the magnetic moment in the AFM phase~\cite{Ishibashi,Mazin2}: The calculations yield a strongly ordered AFM state with the magnetic moment of $\sim2$~$\mu_\mathrm{B}$, which is larger than the experimental ones of $<1$~$\mu_\mathrm{B}$~\cite{Cruz}.
To this discrepancy in magnetic moment, several interpretations have been provided, such as the presence of domain motions~\cite{Mazin3}, dynamical spin fluctuations~\cite{Singh}, and the requirement of negative $U$~\cite{nakamura}.
Still unknown is whether the weak order model is appropriate ---though we know the strong order model seems inappropriate for the iron pnictides--- since the properties of weakly ordered states, apart from the small magnetic moment, have not been much studied yet.

The magnetic order strength affects not only the magnetic excitations but also the charge excitations because of the presence of gap (not a full but a partial gap).
In fact, the in-plane optical conductivity for BaFe$_2$As$_2$ and SrFe$_2$As$_2$~\cite{Hu} as well as EuFe$_2$As$_2$~\cite{Wu} has shown significant change of its spectra with reducing temperature across $T_N$.
A new multi-peak structure appears at around 0.05-0.18~eV depending on material, while the spectra above $\sim 0.6$~eV is almost unchanged.
A similar peak structure has also been observed in polycrystalline LaFeAsO~\cite{Boris}.
It has not yet been clarified whether such a spectral change is consistent with the emergence of the weak magnetic order.
This is crucial for modeling the AFM phase as well as the superconducting phases of the iron pnictides.

In this Letter, we discuss the excitation properties of the weak order model.
Investigating the optical conductivity arising from interband transitions for weakly ordered states, we verify that the weak order model reproduces the experiments and depict the system well.
In addition, we predict the field dependence of the interlayer magnetoresistance in this system.

\textit{Mean-field five-band model.}---
Considering an Fe square lattice, we start with the Hamiltonian for a $d$-electron system $H = H_0 +H_I$.
Here,
\begin{eqnarray}
\hspace{-0.7cm}H_0=\sum_{\mathbf{k},\mu,\nu,\s}
\[
\sum_{\Delta} t
(\Delta_x,\Delta_y;\mu,\nu)
\e^{\i \mathbf{k}\cdot\Delta}
+\epsilon_{\mu}\delta_{\mu,\nu}\]
c_{\mathbf{k}\mu\s}^\dagger c_{\mathbf{k} \nu\s}
\end{eqnarray}
is the five-band hopping Hamiltonian, where $c_{\mathbf{k}\nu\s}^\dagger$ creates an electron with a wave vector $\mathbf{k}$ and a spin $\s$ at an orbital $\mu$, the hopping energy $t(\Delta_x,\Delta_y;\mu,\nu)$ is given from Ref.~\cite{Kuroki}, and $\Delta=(\Delta_x,\Delta_y)$.
$H_I$ is the interaction Hamiltonian~\cite{Oles}:
\begin{eqnarray}
H_I&=& U\sum_{i,\mu}
n_{i \mu\up}n_{i \mu\dn}
+\frac{2U-5J}{4}
\sum_{i,\mu\neq\nu,\s,\s'}
n_{i \mu\s}n_{i \nu\s'}
\nonumber\\
&&+J\sum_{i,\mu\neq\nu}
(c_{i \mu\up}^\dagger c_{i \nu\up}
c_{i \mu\dn}^\dagger c_{i \nu\dn}
-\mathbf{S}_{i\mu}\cdot\mathbf{S}_{i\nu}),
\end{eqnarray}
where $i$ is the Fe-site index; $U$, the intra-orbital Coulomb interaction; $J$, the Hund coupling.
The number and the spin operators are defined as
$n_{i \mu\s}=c_{i\mu\s}^\dagger c_{i\mu\s}$ and $\mathbf{S}_{i\mu}=\frac{1}{2}\sum_{\tau,\tau'}c_{i\mu\tau}^\dagger \hat{\s}_{\tau\tau'} c_{i\mu\tau'}$, respectively, with $\hat{\s}$ the Pauli spin matrix.
We set the Fe-Fe bond length $a_0$ to be unity and $x$ and $y$ to be along the nearest Fe-Fe bond directions.

To obtain a model of the ordered state in iron pnictides, we solve mean-field equations selfconsistently in the same manner as in Ref.~\cite{Ran}.
Considering the magnetic ordered states with the spin-density-wave (SDW) ordering vector $\mathbf{Q}=(\pi,0)$, we take the order parameters such as
$
\<n_{\mathbf{Q}\,\mu\nu\s}\>=\frac{1}{N}\sum_{\mathbf{k}}
\<c_{\mathbf{k}+\mathbf{Q}\,\mu \s}^\dagger c_{\mathbf{k}\,\nu \s}\>
$
with $N$ the number of $\mathbf{k}$ points in the first Brillouin zone of the five-band paramagnetic system.
We obtain the quasiparticle state
$
\gamma^\dagger_{\mathbf{k}\epsilon\s} =\sum_{l=0,1}\sum_{\mu} \psi_{\mu\epsilon\s}(\mathbf{k}+l\mathbf{Q}) c^\dagger_{\mathbf{k}+l\mathbf{Q}\mu\s}
$
with the energy $E_{\mathbf{k\epsilon\s}}$.

\begin{figure}[t]
\begin{center}
\includegraphics[width = 0.9\linewidth]{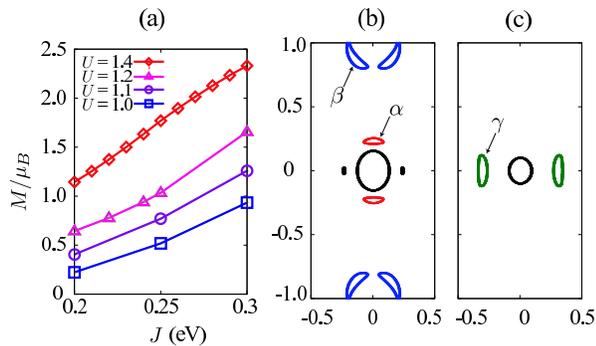}
\caption{(Color online) (a) The magnetic moment per Fe site for different parameter sets: $U=1.0$~eV (squares), 1.1~eV (circles), 1.2~eV (triangles), and 1.4~eV (diamonds).
Fermi surfaces are plotted for the weak order case (b) and the strong order case (c) in the magnetic Brillouin zone of the five-band model ($-\frac{1}{2}<\frac{k_x}{\pi}<\frac{1}{2}$, $-1<\frac{k_y}{\pi}<1$).
The labeled Fermi surfaces ($\alpha$, $\beta$, and $\gamma$) contribute to the interlayer magnetoresistance (see text).
} \label{fig:mag}
\end{center}
\end{figure}

In our calculation, we change the parameters $U$ and $J$ to control the order strength expressed by the magnetic moment
$
M=\sum_{\mu}\<n_{\mathbf{Q}\,\mu\mu\up} -n_{\mathbf{Q}\,\mu\mu\dn}\> \mu_\mathrm{B}
$
with $\mu_B$ the Bohr magneton.
The calculated magnetic mometns for several parameter sets are shown in Fig.~\ref{fig:mag}(a).
The computations are performed on a system with $N=400\times400$.

For the subsequent arguments, we take the following two parameter sets, which give the magnetic moments $M=0.4\mu_\mathrm{B}$ and $2.3\mu_\mathrm{B}$ [Fig.~\ref{fig:mag} (a)], as models of weakly and strongly ordered states, respectively: $U=1.1$~eV, $J=0.2$~eV; and $U=1.4$~eV, $J=0.3$~eV.
The former agrees well with the experiments~\cite{Cruz} on the magnetic moment.
The latter agrees well with the first-principles calculations not only on the magnetic moment but also on the Fermi surface and the partial density of states~\cite{Shimojima}.
For the parameter sets that give similar magnetic moments, we obtain results with similar features to those discussed in this Letter.

The magnetic ordering is achieved with a SDW gap; this is not a full gap, and the system stays metallic.
The Fermi surfaces are drawn in Figs.~\ref{fig:mag}(b) and (c), the data of which are used to calculate the interlayer magnetoresistance below.
For other parameter sets, the states with a similar magnetic moment possess a similar Fermi surface.
In all cases, the circular Fermi surface at $(0,0)$ is holelike and others are electronlike.
There are two circular Fermi surfaces around (0,0) in the paramagnetic case~\cite{Kuroki}; the SDW gap partially opens along the outer (electronlike) one owing to the Fermi surface nesting, and small segments remain at $\alpha$ or $\gamma$.

In the weak order model, the states on the hole pockets at $(0,0)$ have more the $d_{yz}$ and the $d_{zx}$ characters than the other orbitals; especially, the $d_{yz}$ ($d_{zx}$) character is dominant near $k_x=0$ ($k_y=0$).
In the electron pocket $\alpha$ in Fig.~\ref{fig:mag}(b), the $d_{zx}$ character is dominant along the flat region on the side near $(0,0)$; the $d_{yz}$ character is dominant, on the other side.
Most of the states on $\beta$ in Fig.~\ref{fig:mag}(b) have the $d_{xy}$ character, and the other states on $\beta$ show the $d_{xy}$ and $d_{zx}$ characters.
The Fermi surfaces observed in a recent angle-resolved photoemission spectroscopy study on BaFe$_2$As$_2$~\cite{Yi} are similar to $\alpha$ and $\beta$.

In the strong order case, the hole pocket at (0,0) has the $d_{zx}$ character dominantly all around it.
The states on $\gamma$ in Fig.~\ref{fig:mag}(c) possess a mixed character of several orbitals:
The states on the $(0,0)$ side show rather the $d_{x^2-y^2}$ character; on the other side, the $d_{zx}$ and $d_{xy}$ characters.

\textit{Optical conductivity.}---
The optical conductivity experiments on iron pnictides have revealed the excitation structure in the magnetically ordered state, whose characteristics become clear by comparing its spectra with those in the paramagnetic state at $T=300$~K.
The paramagnetic state shows a broad peak at around 0.5~eV, and no peaks below this apart from the Drude peak~\cite{Hu}.
Compared with the paramagnetic state, the magnetically ordered state shows the following features~\cite{Hu}:
(i) an emergence of the excitations related to the SDW gap at around 0.1~eV, (ii) a slight energy shift of the 0.5~eV structure, and (iii) almost the same structure above 0.5~eV.
We use these features (i)-(iii) as criteria of judgement in evaluating the model of the AFM state in Fe pnictide system.

Now we calculate the optical conductivity for the weak order model to examine whether the weakly ordered state can model the AFM state.
We also discuss the strong order model from the same view point.
The real part of the optical conductivity is obtained as follows:
\begin{eqnarray}
&&\hspace{-1cm}\Re{\s_{\alpha\beta}(\omega)}=\frac{-\pi(e/\hbar)^2}{4N\omega}
\sum_{\s,\mathbf{k},\epsilon,\epsilon'}
\[f(E_{\mathbf{k}\epsilon\s})-f(E_{\mathbf{k}\epsilon'\s})\]
\nonumber\\
&&\hspace{1.5cm}\times
\zeta^{(\alpha)}_{\mathbf{k}\epsilon\epsilon'\s}
\[\zeta^{(\beta)}_{\mathbf{k}\epsilon\epsilon'\s}\]^*
\delta(E_{\mathbf{k}\epsilon\s}-E_{\mathbf{k}\epsilon'\s}-\omega),
\end{eqnarray}
where $f$ is the Fermi distribution function, $e$ is the elementary charge, and $\zeta^{(\alpha)}_{\mathbf{k}\epsilon\epsilon'\s}$ is defined as
\begin{eqnarray}
&&\hspace{-.5cm}\zeta^{(\alpha)}_{\mathbf{k}\epsilon\epsilon'\s}
=
\sum_{\Delta,\mu,\nu}
\Delta^{(\alpha)}
t(\Delta;\mu,\nu)
\Big(\sum_{l=0,1}
\psi_{\mu\epsilon\s}(\mathbf{k}+l\mathbf{Q}) \psi^*_{\nu\epsilon'\s}(\mathbf{k}+l\mathbf{Q})
\nonumber\\
&&\hspace{0cm}
\big\{ (1-s_{\mu\nu})
\cos[(\mathbf{k}+l\mathbf{Q})\cdot\Delta]
- \i(1+s_{\mu\nu})
\sin[(\mathbf{k}+l\mathbf{Q})\cdot\Delta]
\big\}\Big)
\nonumber\\
\end{eqnarray}
with $\Delta^{(\alpha)}$ the $\alpha$ component of the vector $\Delta$ and $s_{\mu\nu}$ ($=\pm1$) defined as
$t(\Delta_{ji};\mu,\nu)= s_{\mu\nu}t(\Delta_{ij};\mu,\nu)$.
We focus on the interband transitions but no Drude component.

\begin{figure}[t]
\begin{center}
\includegraphics[width = 1.0\linewidth]{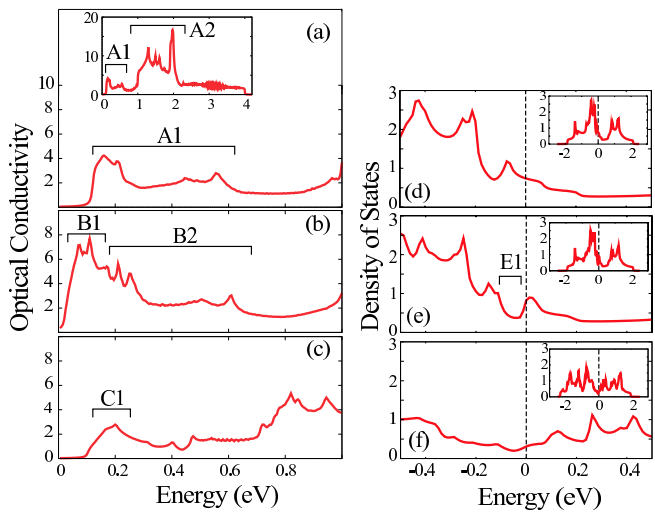}
\caption{(Color online) Optical conductivity and density of states for the paramagnetic [(a) and (d)], the weakly ordered [(b) and (e)], and the strongly ordered [(c) and (f)] cases.
The inset in (a) shows the whole excitation spectra of interband transitions.
Some characteristic excitation structures and a gap are labeled such as A1, A2, etc.
The optical conductivity is normalized by $\frac{(e/\hbar)^2}{2}$.
The insets in (d)-(f) show the whole structure.
} \label{fig:optcond}
\end{center}
\end{figure}
The calculated optical conductivities for the direction of $\alpha=\beta=(1,1)$ are plotted in Figs.~\ref{fig:optcond}(a)-(c).
Within our calculations, the temperature effect is only to broaden the excitation peaks through the Fermi distribution function $f$, so that we set $T=0$ and neglect the effect in the following calculations.
(We do not argue the experimentally observed gap-like behavior above $T_N$ below around $0.6$~eV, the so-called psuedogap feature~\cite{Hu,Wu,Boris,Ikeda};
this is possibly due to the spin-fluctuation effects~\cite{Prelovsek} beyond our mean-field calculation.)

First, we discuss the spectra in the paramagnetic case [Fig.~\ref{fig:optcond}(a)].
There are two excitation structures: One lies in the energy range 0.1-0.6~eV (labeled as A1); the other, above 1.0~eV (A2 in the inset).
The double broad-peak structure below and above 1.0~eV is also observed in the experiments on arsenic compounds~\cite{Hu} and phosphorus ones~\cite{Qazilbash}.
However, the A1 and A2 have a fine structure [Fig.~\ref{fig:optcond}(a)]; the appearance of the structure differs from the single-broad-peak structure in the experiments around 0.5 eV.
In addition, the intensity ratio of A2 to A1 is much higher than that in the experiments.
These differences are associated with effects not included in our calculation, such as spin-fluctuation effects.
Except for these details, the broad aspect seems consistent with the experiments.

Next, we argue the weak order case [Fig.~\ref{fig:optcond}(b)], comparing with the paramagnetic one.
In the weakly ordered case, the excitations related to the SDW gap appear around 0.1~eV (B1).
The other parts of the excitation spectra are almost the same as that in the paramagnetic case, except for the small shift from A1 to B2.
These features agree well with the experimental results on the criteria (i)-(iii).
It is natural that these three features appear because the weakly ordered state has an electronic structure similar to that of the paramagnetic one [see Figs.~\ref{fig:optcond}(d) and (e)], and the difference lies only near the Fermi level, where the SDW partial gap opens [indicated as E1 in Figs.~\ref{fig:optcond}(e)].
The interband transitions that contribute the B1 excitations occur at the SDW gap near $(\pi/4,0)$, which mainly involve the orbital transition from $d_{xy}$ to $d_{yz}$.

In the strongly ordered case, on the other hand, the density of states and the excitation spectra differ from the paramagnetic case [see Figs.~\ref{fig:optcond}(c) and (f)].
The excitations related to the SDW gap appear around 0.2~eV [C1 in Fig.~\ref{fig:optcond}(c)], whose energy is inconsistent with the criterion (i).
Other excitation peaks lie above 0.5~eV, whose energies appear to be shifted up by $\geq0.5$~eV as compared with A1 and A2 (the energy region $\geq1.0$~eV not shown here).
This disagrees with the criteria (ii) and (iii).
Therefore, as is already suggested from the magnetic moment study, the strong order model is certainly not a valid model in terms of the excitation properties as well.

Here, we mention three dimensional effects.
The first-principles calculations show three dimensional Fermi surfaces especially in the systems such as BaFe$_2$As$_2$ and suggest those effects be important.
We confirmed that the above behavior of the optical conductivity, however, is unchanged even in the presence of the $d$-$d$ hopping in the $z$ direction (not shown here).

Hence, our optical conductivity calculations suggest that the weakly ordered state be a valid model of the ordered state in iron pnictides.

\textit{Interlayer magnetoresistance.}---
For a further experiment to examine the weak order model, we propose an interlayer magnetoresistance experiment.
Below, calculating the interlayer magnetoresistance, we provide an aspect of its magnetic-field and azimuthal-angle dependence for the benefit of experimental research.
The interlayer resistance at zero temperature $\rho_{zz}(\phi)$ under a magnetic field $\mathbf{B}=B(\cos(\phi),\sin(\phi),0)$ parallel to the layer is given by~\cite{morinari}
\begin{eqnarray}
\frac{1}{\rho_{zz}(\phi)}
=\frac{e^2}{2\pi^2}
\(\frac{t_c a_c}{\hbar}\)N_z
\sum_{\epsilon}
\int_{E_{\mathbf{k}\epsilon\s}=E_F}
\frac{d\ell_\mathbf{k}}{|\mathbf{v}_{\mathbf{k}\epsilon\s}|}
\frac{\Gamma}{\Xi}
\label{eq:mr}
\end{eqnarray}
with $t_c$ the interlayer tunneling, $c$ the speed of light in vacuum, and $a_c$ the interlayer distance, where
$\Xi=\left| \frac{ea_c}{c}
\(\mathbf{v}_{\mathbf{k}\epsilon\s}\times\mathbf{B}\)
\right|^2+\Gamma^2$, and impurity scattering effects upon interlayer hoppings are included by the constant parameter $\Gamma=\hbar/(2\tau)$ with $\tau$ the scattering time.
Hereafter, we set $\Gamma=2.0\times10^{-4}$~eV.

\begin{figure}[t]
\begin{center}
\includegraphics[width = 1.0\linewidth]{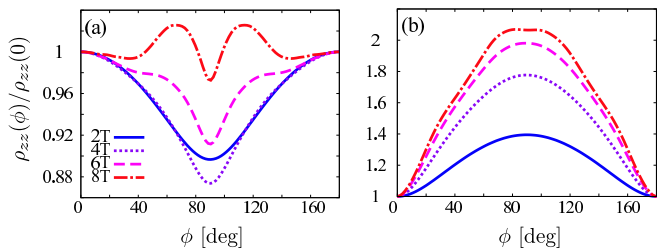}
\caption{(Color online) The azimuthal-angle dependence of the interlayer magnetoresistances for the weak order model (a) and the strong order (b) under magnetic fields $B=2$~T (solid), 4~T (dotted), 6~T (dashed), and 8~T (dash-dotted).
} \label{fig:mr}
\end{center}
\end{figure}

In Fig.~\ref{fig:mr}, the calculated interlayer magnetoresistances are plotted.
In the weak order case [Fig.~\ref{fig:mr}(a)], as $B$ increasing from 2 to 6~T, the local minimal value of $\rho_{zz}(\phi)/\rho_{zz}(0)$ decreases first and then increases.
Furthermore, at 8~T, $\rho_{zz}(\phi)/\rho_{zz}(0)$ shows a double-peak structure.

Here, we interpret the formula (\ref{eq:mr}) to explain the above features of $\rho_{zz}(\phi)/\rho_{zz}(0)$.
It follows from the formula (\ref{eq:mr}) that a Fermi surface with a flat region leads to an oscillating behavior in the $\phi$ dependence of $\rho_{zz}(\phi)/\rho_{zz}(0)$~\cite{morinari}.
Contrary, a circular Fermi surface hardly contributes to the $\phi$ dependence.
In fact, the $\phi$ dependence mainly arises from the two Fermi surfaces $\alpha$ and $\beta$ in Fig.~\ref{fig:mag}(b) (and their symmetric counterparts); in addition, each shows a different $\phi$ dependence as well as a different $B$ dependence.

For a small $B$ ($\ll2$~T), the formula (\ref{eq:mr}) is almost independent of $\phi$ (note that $\Xi^{-1}\approx\Gamma^{2}$), and the $\phi$ dependence of $\rho_{zz}$ is negligible.
The $\phi$ dependence emerges when $B$ becomes larger than a threshold, which depends on the Fermi velocities and $\Gamma$ ---the larger the Fermi velocities, the lower the threshold.
By investigating the Fermi velocity, we find the threshold for $\alpha$ is lower than $\beta$.

For $B=2$~T and 4~T, the $\alpha$ contribution is dominant and tends to generate a valley in $\rho_{zz}(\phi)/\rho_{zz}(0)$ at $\phi=90$, owing to the flat segment of $\alpha$.
The $\beta$ contribution, on the other hand, emerges for $B\ge6$~T and tends to develop a hump at $\phi=90$ owing to the flat segment at around $(\pi,0.2\pi)$; this suppresses the valley structure there.
Hence, the crossover of the dominant contribution of the Fermi surface results in the non-monotonic $B$ dependence.
Also, the double-peak structure at 8 T in Fig.~\ref{fig:mr}(a) is attributed to the total contribution of $\alpha$ and $\beta$.
For larger $B$, the hump component due to $\beta$  at $\phi=90$ develops more since the flat segment of $\beta$ is longer than that of $\alpha$.

For the strong order model [Fig.~\ref{fig:mr}(b)], in contrast, the $\phi$ dependence at $B=2$~T is 90-degree shifted from that for the weak order model.
In addition, we find a clear difference in the $B$ dependence:
$\rho_{zz}(\phi)/\rho_{zz}(0)$ changes monotonically unlike the weak order case.
This is because only one type of Fermi surface ---$\gamma$ in Fig.~\ref{fig:mag}(c)--- contributes to it in the strong order model.

Hence, we predict a non-monotonic $B$ dependence of the interlayer magnetoresistance reflecting the presence of the Fermi surface $\beta$ in the weak order model, rather than a monotonic one in the strong order model.
We add that, with a sufficiently weak scattering, the double-peak structure at $\sim8$~T would be observable ---this structure may appear as a dip structure for larger $B$.
These two features (the non-monotonic behavior and the double-peak structure) are the crucial consequence of the multi Fermi surfaces ($\alpha$ and $\beta$).
Although the above argument is based on the calculations with the scattering rate that is the same for each band and isotropic, we can discuss it from the same point of view even for the case of different scattering rates.
From calculations with different scattering rates for each band (not shown here), we verified that at least one of the two features arising from the multi Fermi surface appears in the weak order case.
In addition, we examined the case of the anisotropic scattering rate~\cite{Hussey}.
The anisotropy causes ripples on the angle dependence of the interlayer magnetoresistance; still, the features arising from the multi Fermi surface is robust against the presence of the anisotropy.

\textit{Conclusions.}---
We have investigated the weakly ordered state as a model of the striped AFM state in the parent compounds of iron-based superconductors.
Compared with the experiments of the optical conductivity, our calculations suggest that the weak order model well reproduce the characteristics of the experimental results in terms of the criteria (i)-(iii).
We have also verified that the strong order model, which corresponds to the model obtained from the first-principles calculations, does not reproduce the experiments.
From these calculations, we conclude that the weak order model is a valid model of this system.
In addition, we have calculated the interlayer magnetoresistance for the models.
From the calculation results, we predict that the interlayer resistance changes non-monotonically as the magnetic field increasing up to 8 T and a double-peak (or a dip) structure under large magnetic fields.
Further experiments on interlayer magnetoresistance are desired to validate our conclusion of the weak order model.

This work was supported by the Grant-in-Aid for Scientific Research from the Ministry of Education, Culture, Sports, Science and Technology of Japan; the Global COE Program ``The Next Generation of Physics, Spun from University and Emergence"; and Yukawa Institutional Program for Quark-Hadron Science at YITP.
Numerical computation in this work was carried out at the Yukawa Institute Computer Facility.


\end{document}